\documentclass{IEEEtran}
\usepackage{cite}
\usepackage{amsmath,amssymb,amsfonts}
\usepackage{graphicx}
\usepackage{textcomp,nicefrac}
\usepackage{lineno}
\usepackage{xcolor}
\usepackage{url}

\def\BibTeX{{\rm B\kern-.05em{\sc i\kern-.025em b}\kern-.08em
T\kern-.1667em\lower.7ex\hbox{E}\kern-.125emX}}
\begin{document}
\title{Study of using machine learning for level 1 trigger decision in JUNO experiment}
\author{{Barbara Clerbaux$^{1}$, Pierre-Alexandre Petitjean$^{1}$, Yu Xu$^{2,3}$, \underline{Yifan Yang}$^{1}$}
\thanks{Submitted for review on October 31, 2020 \\
$^{1}$Inter-university Institute for High Energies, Universit\'e libre de Bruxelles (ULB)\\
$^2$ Forschungszentrum J\"ulich IKP  \\
$^3$ III. Physikalisches Institut B, RWTH Aachen University}}

\maketitle

\begin{abstract}
A study on the use of a machine learning algorithm for the level 1 trigger decision in the JUNO experiment is presented. JUNO is a medium baseline neutrino experiment in construction in China, with the main goal of determining the neutrino mass hierarchy. A large liquid scintillator (LS) volume will detect the electron antineutrinos issued from nuclear reactors.
The LS detector is instrumented by around 20000 large photomultiplier tubes.
The hit information from each PMT will be collected into a center trigger unit for the level 1 trigger decision. 
The current trigger algorithm used to select a neutrino signal event is based on a fast vertex reconstruction.
We propose to study an alternative level 1 (L1) trigger in order to achieve a similar performance as the vertex fitting trigger but with less logic resources by using firmware implemented machine learning model at the L1 trigger level.
We treat the trigger decision as a classification problem and train a Multi-Layer Perceptron (MLP) model to distinguish the signal events with an energy higher than a certain threshold from noise events.
We use JUNO software to generate datasets which include 100K physics events with noise and 100K pure noise events coming from PMT dark noise.
For events with energy higher than 100 keV, the L1 trigger bosed on the converged MLP model can achieve an  efficiency higher than 99\%.
After the training performed on simulations, we successfully  implemented the trained model into a Kintex 7 FPGA.
We present the technical details of the neural network development and training, as well as its implementation in the hardware with the FPGA programming.
Finally the performance of the L1 trigger MLP implementation is discussed. 
\end{abstract}


\section{The JUNO experiment}
\begin{figure}[t!]
\centering
\includegraphics[width=3.5in]{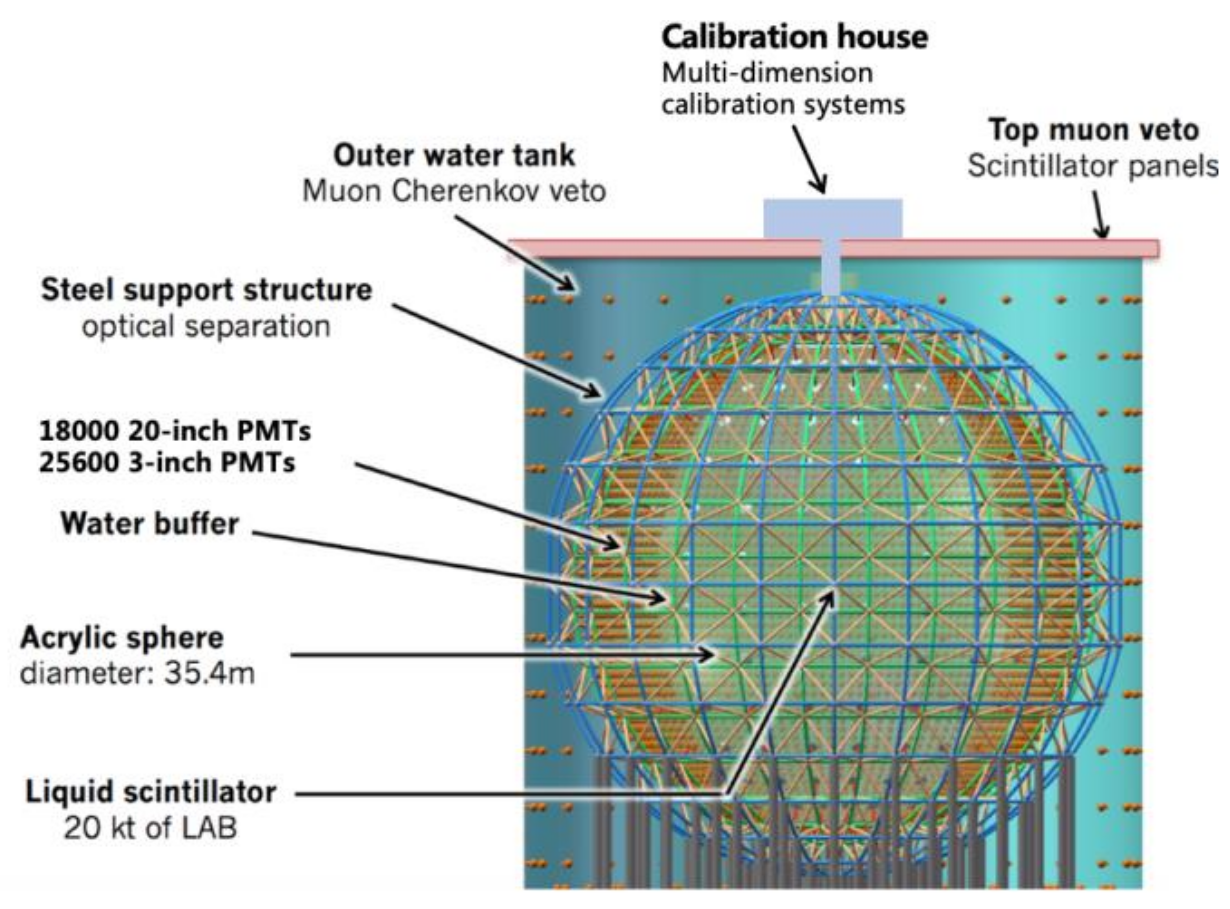}
\caption{The JUNO detector with the central acrylic sphere, the two veto systems and the calibration part.}
\label{figure:detector}
\end{figure}

The Jiangmen Underground Neutrino observatory (JUNO) experiment \cite{Djurcic} uses a large liquid scintillator detector aiming at measuring electron antineutrinos issued from nuclear reactors at a distance of 53 km. The main goal to determine the neutrino mass hierarchy, after 6 years of data taking \cite{JUNO_phys}.
The detector will be located at 700 m underground and will consists of 20 ktons of liquid scintillator contained in a 35 m diameter acrylic sphere, instrumented by 18000 20-inch photomultiplier tubes (PMT) and 25600 3-inch PMTs.
Two vetoes are foreseen to reduce the different backgrounds. A 20 ktons ultrapure water Cerenkov pool around the central detector instrumented by 2000 20-inch PMTs will tag events coming from outside the neutrino target. It will also act as a passive shielding for neutrons and gammas.
In addition, a muon tracker will be installed on top of the detector (top muon veto) in order to tag cosmic muons and validate the muon track reconstruction. A schematic view of the detector is presented in figure \ref{figure:detector}.

\section{Electronics readout system}
The JUNO electronics system will have to cope with signals from 18000 large (20-inch) PMTs and 25600 small (3-inch) PMTs of the central detector as well as 2000 PMTs installed in the surrounding water pool. One of the innovative aspects of JUNO is its electronics and readout concept \cite{Djurcic}.
With JUNO, a flexible approach is chosen: every 3 PMTs will have their own “intelligence” as self-monitoring (currents, voltages, temperatures) and it will have the possibility to implement various data processing algorithms (for example, trigger request generation and data compression).
The JUNO electronics system can be separated into mainly two parts: (i) the front-end electronics system, performing the analog signal processing (the underwater electronics) and after about 100 m cables, (ii) the back-end electronics system, sitting outside water, consisting of the data acquisition (DAQ) and the trigger. 
The main challenge of the whole electronics system is the very strict criteria on reliability: a maximum of 0.5\% failure over 6 years for the PMT full readout chain, as well as the large data transfer of 1.25 Gb/s that needs to be delivered over 2 times 100 m Ethernet cables. A detailed description of the full electronics chain of JUNO, from the PMTs front-end to the trigger and DAQ is presented in \cite{Bellato}.\\

\begin{figure}[t!]
\begin{center}
\centering
\includegraphics[width=3.2in]{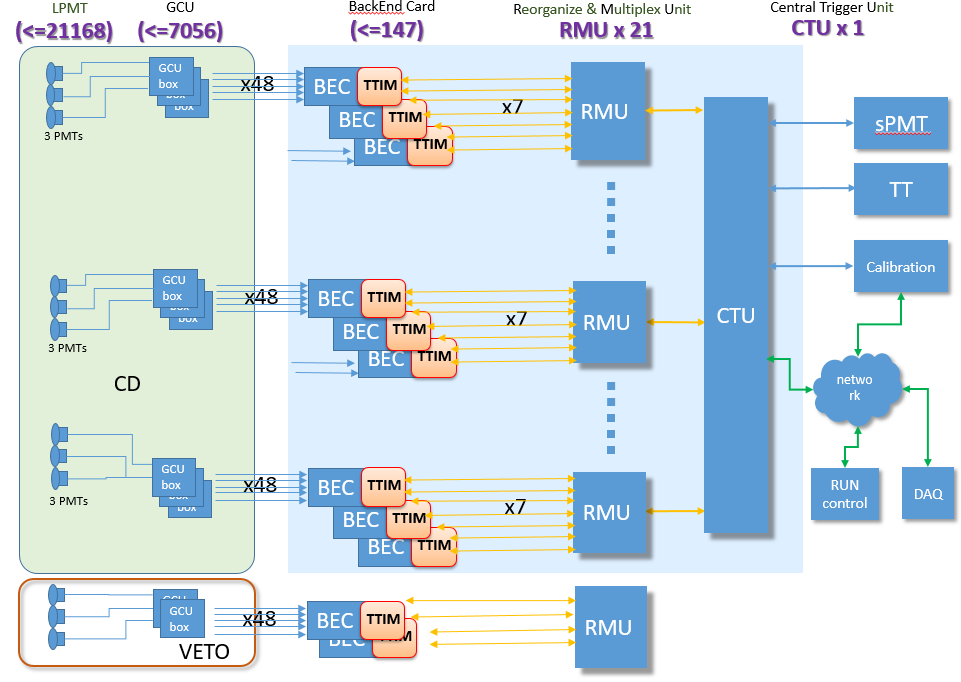}
\caption{The electronics readout system of the JUNO experiment. The underwater electronics (PMT, Global Control Unit ) is presented in the left (green) box, and the dry electronics (Back-End Card , Reorganize \& multiplex Unit  and the Central Trigger Unit ) in the right (blue) box.}
\label{figure:trigger}
\end{center}
\end{figure}

For the front-end electronics, the global control unit (GCU) \cite{GCU} will digitize the incoming analog signals with custom designed high speed ADU (analog to digital converter unit). 
The GCU will store the signals in a large local memory under the control of the FPGA (Field-Programmable Gate Array) waiting for the trigger decision and sending out possible event data as well as trigger requests to the outside-water system. 
A back-end card will be used as a concentrator and each of the incoming trigger request signals will pass through an equalizer for compensating the attenuation due to the long cables \cite{yang2018design}.
An FPGA mezzanine card (called TTIM, for trigger timing interface mezzanine) will collect all differential trigger request signals, aligning them with certain trigger count. Then it will make a sum, and send the result to the next stage (RMU, for Reorganize \& Multiplex Unit) and finally to the Center Trigger Unit (CTU) over optical fibers.
A schematic view of the electronics readout system is presented in figure \ref{figure:trigger}.

\section{Trigger algorithms}
\begin{figure}[t!]
\begin{center}
\centering
\includegraphics[width=3.0in]{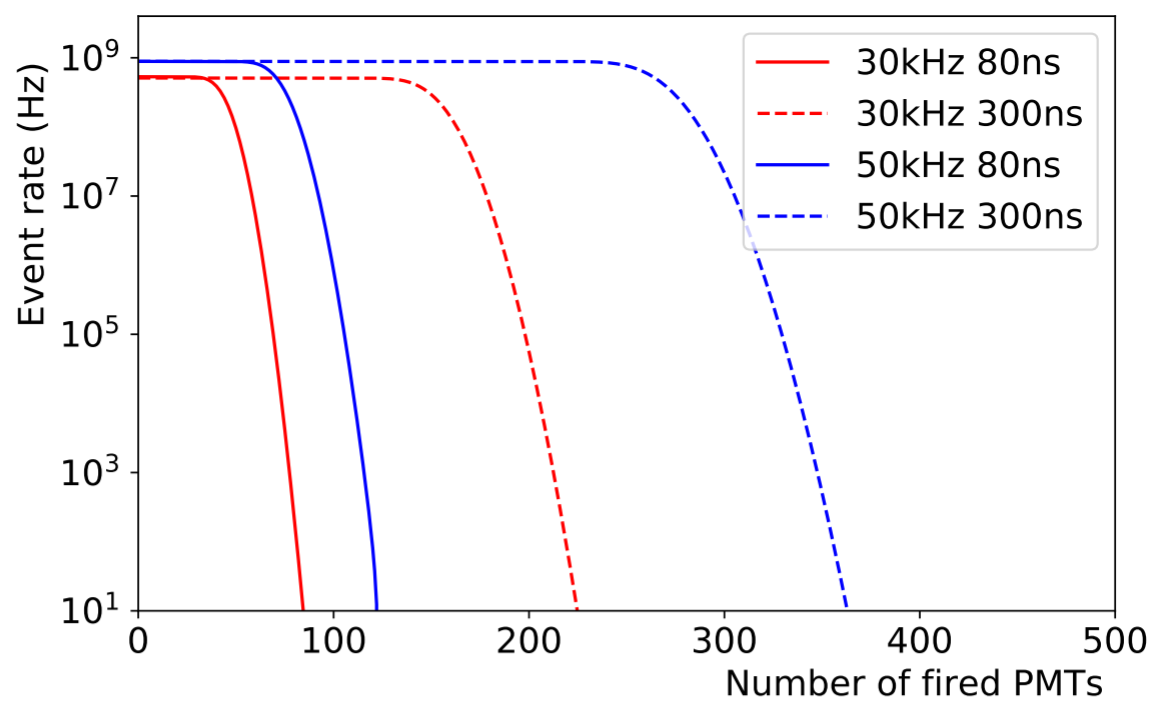}
\caption{Dark noise event rate as a function of the number of fired PMTs. The full (dotted) lines use a time window of 80 ns (300 ns). Two dark noise frequencies (30 KHz and 50 KHz) are also considered  \cite{globtrig}.}
\label{figure:motiv}
\end{center}
\end{figure}

Two possible trigger algorithms have been discussed in JUNO and are detailed in reference \cite{Fang2020}. The first and simpler one, called the "multiplicity trigger", counts the number of fired PMTs in a 300 ns timing window. Figure \ref{figure:motiv} presents the dark noise event rate as a function of the number of fired PMTs for two different dark noise frequencies (30 KHz and 50 kHz) and for 2 different time windows (80 ns and 300 ns). It can be observed that a high trigger threshold on the number of PMTs fired has to be set in order to reject the coincidences originated from the PMT dark noise, which will sacrifice the recognition of low energy events.

\begin{figure}[t!]
\begin{center}
\centering
\includegraphics[width=3.0in]{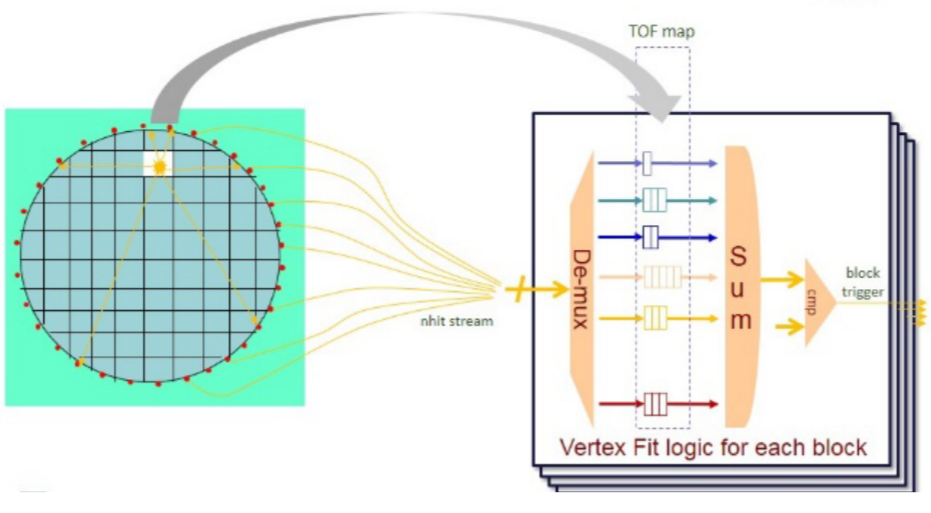}
\caption{Vertex fitting concept in JUNO. On the left side, an example of an event happening in the upper part of the central detector, and firing several PMTs. On the right side, the schematic view of the vertex fit logic, with a correction of the various times of flight for each PMT, before the sum is performed and send to the trigger bloc.}
\label{figure:vertexfitting}
\end{center}
\end{figure}

In order to reduce the impact from the PMT dark noise, a sophisticated trigger scheme called vertex fitting is proposed in references \cite{Fang2020} and \cite{globtrig}. Unlike the "multiplicity trigger", the vertex fitting method tests the PMT information with all possible positions and finds the most likely one among them. 
In order to estimate the vertex position, it is needed to extract the charge information form each hit. The whole detector volume is divided into a certain number of blocks. 
By correcting the time of flight for each PMT, it is possible to decrease the timing window to 80 ns while reducing the effect from the PMT dark noise even for 0.1 MeV physical events. 
The schematic view of the vertex fitting concept is shown in figure \ref{figure:vertexfitting}.
The candidate hardware for implementing the vertex fitting is a Virtex 7 FPGA.

The goal of the study presented in this article is to find an alternative way, using a  machine learning model, to distinguish between signal events at low energy (as low as 100 keV) and background events from dark noise, while keeping the algorithm lightweight enough to be implemented in a Kintex 7 FPGA.
 
 We describe our study in the following 3 chapters. We start with a description of the signal and noise event simulations, followed by the MLP model training. Finally we present the firmware implementation.

\section{Simulation of signal and noise events}

\begin{figure}[ht!]
\centering
\includegraphics[width=3.5in]{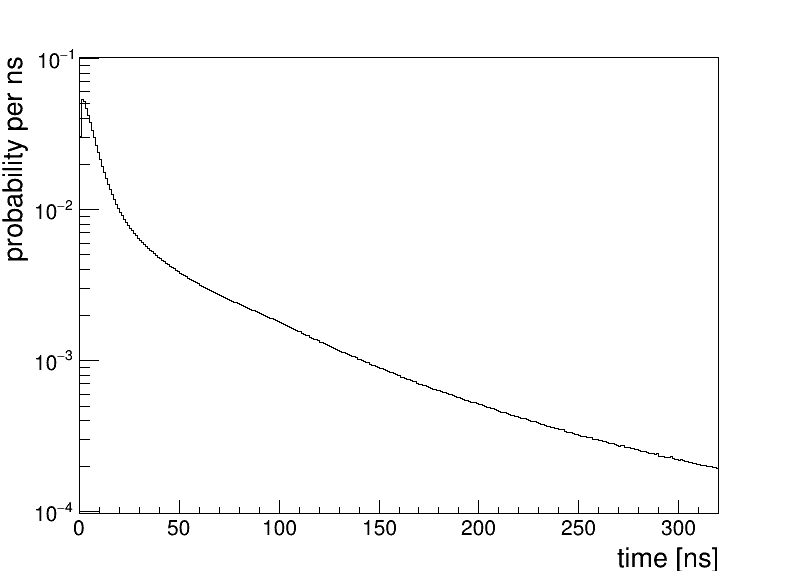} 
\caption{The time profile of one signal event. Here we use alpha particles as an example of signal event.}
\label{figure:DS}
\end{figure}

In the JUNO experiment, when a signal event happens, the kinetic energy of the final state particles can be transferred to photons that are detected by the PMTs. The initial information of the events is thus the observed hits distributed in a time window. In JUNO, the signal events include neutrino events, radioactive decay events, and cosmic ray events, and the hits are always distributed in the shape of events' time profile in the LS as shown in figure \ref{figure:DS}. 

However, we cannot collect only the "pure" number of hits from signal events, since these hits are always contained in the dark noise coming from the PMT electronics.
Although the dark noise hits are distributed uniformly in the time window, the number of dark hits in one time window follows a Poisson distribution, and we cannot know the accurate number of dark hits in the time window. Therefore, when the number of  detected hits is close to the average level of dark noise, it is hard to say whether the time window contains a real signal event. This is specially the case for the low energy signal events. In that case, it is a challenging topic to identify signal event hits from pure dark noise hits.  

\begin{figure}[ht!]
\centering
\includegraphics[width=3.5in]{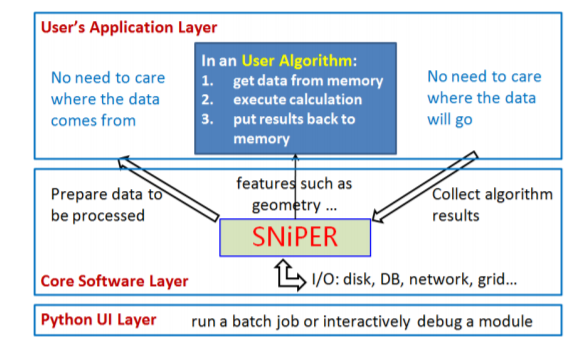} 
\caption{An overview of SNiPER structure with three layers : the Python UI layer, the core software layer and the user's application layer \cite{Zou_2015}.}
\label{figure:Sniper}
\end{figure}

In this study, we want to identify two kinds of data: the pure dark noise events and the signal events with dark noise. For the signal events, we prepared simulated samples using the SNiPER (Software for Non-collider Physics ExpeRiments) software \cite{Lin_2017}. SNiPER is a unified software platform based on common requirements from both nuclear reactor neutrino experiments and cosmic ray experiments. It is used  by the JUNO and LHAASO (Large High Altitude Air Shower Observatory) experiments \cite{Zou_2015}. The SNiPER framework consists of physics generators, detector simulation and electronics simulation modules. The JUNO detector simulation software is based on Geant4, while the management of the geometry and the user actions are simplified by introducing several interface classes. The software has excellent efficiency and flexibility with an open structure.  An overview of SNiPER structure in 3 layers is shown in figure \ref{figure:Sniper}. 

\begin{figure}[ht!]
\centering
\includegraphics[width=2.5in]{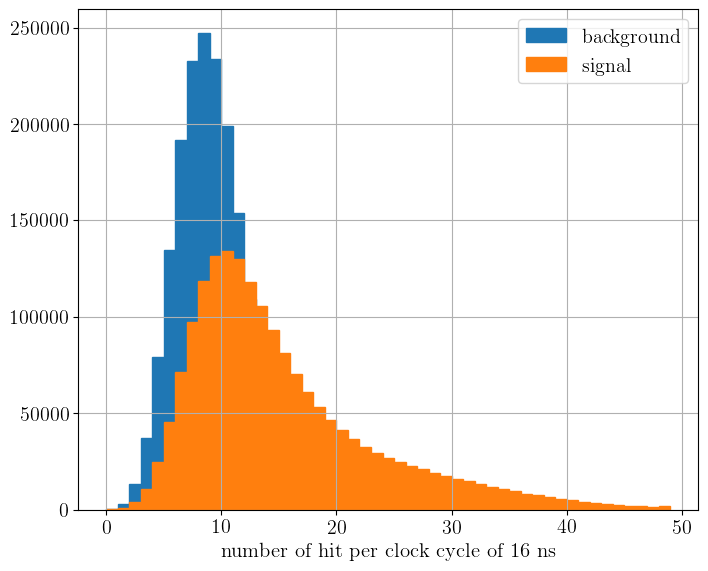} 
\caption{Distribution of fired PMT amount in 16ns for both background events and signal events.}
\label{figure:Distevents}
\end{figure}

In the present study, we generate electron events at fixed energy as the signal events. The electron energy chosen is 100 keV. We generate 100k signal events in total, with positions uniformly distributed in the detector. Then, the dark noise is added to the signal events. For the dark noise simulation, we  generate some hits uniformly distributed in the time window. For the training, we use 100 keV electron events with dark noise as signal events, and pure dark noise events as background events. In JUNO, the light yield is around 1200 p.e./MeV, so there will be in average 120 hits for 100 keV events. We set the number of hits for dark noise in the range of 100 to 240.
Figure \ref{figure:Distevents} shows the distribution of fired PMT amount in one clock cycle for both background events and signal events,it's easy to tell from the plot that two distributions are different while having partly overlap, it's difficult to express the difference with an explicit function.

The data generated by the simulation contains two parts: the label and the inputs for the CTU. The label tells if the event is either a signal event or a dark noise event. The inputs for CTU are the numbers of fired PMTs in a continuously 20 clock cycles, therefore, they are 1$\times$20 arrays, and each element in the array describes the number of fired PMTs in the corresponding clock cycle. The time window is 320 ns for a 62.5 MHz clock, it is long enough to cover any event occurring at any position of the JUNO detector.  For the labels we assign 1 to signal events, and 0 to background events.
Since a trigger decision can be treated as a binary classification problem, what we hope is that the model can assign the correct label from the different inputs given. The inputs are the information which is received by the CTU, and the label is the trigger decision result, either 1 (for a signal event) or 0 (for a dark noise event).

Training the machine learning model requires a big amount of labeled data which we do not have. We thus use MC simulations to simulate specific behaviors of the detector such as the CTU inputs  for different events, the combination of events type. The labeled data are  used to train the machine learning model.

\section{Multilayer perceptron neural network (MLP)}

\begin{figure}[t!]
\centering
\includegraphics[width=3.5in]{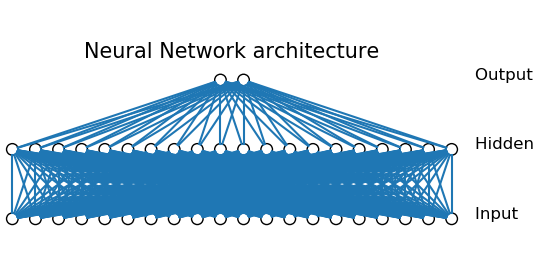}
\caption{Neural network architecture with the input layer, a hidden layer and the output layer.}
\label{figure:NN}
\end{figure}

The structure of the artificial neural network used in the present study is shown in figure \ref{figure:NN}. Each disk represents a neuron and the blue lines correspond to the forward link between the neurons. 

The forward link of a neuron is computed following this formula \cite{Goodfellow-et-al-2016}: 
\begin{equation*}
    Y_k = \displaystyle f \left(\sum_{i=0}^N w_i \times X_i +b \right) 
\end{equation*}
where $Y_k$ is the value of a neuron number k in the layer number $j$ , we do the sum of the over all the neuron of the previous layers (layer number $j-1$) : N is the number of neurons of previous layer,  $X_i$ is the value of the neuron number i of previous layer, $w_i$ is the weight associated to the link between $X_i \to Y_k$, $b$ is the bias of the neuron k in the layer $j$ and $f$ is the activation function applied on the sum. The activation functions used in this paper are the two following one where z is the argument of the function: 
\begin{itemize}
    \item $Relu$ function : $\sigma (z) = \max(0,z)$
    \item $Softmax$ function : $\sigma (z) = \displaystyle \frac{e^{\beta z_k}}{\sum_i e^{\beta z_i}}$
\end{itemize} 

\begin{figure}[ht!]
\centering
\includegraphics[width=4in]{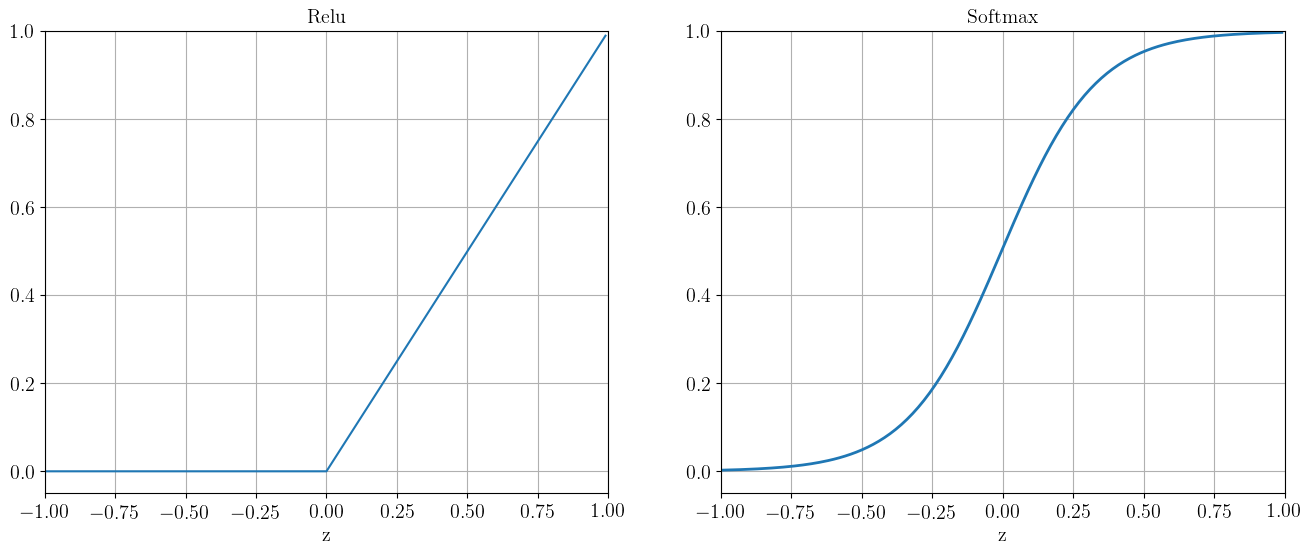}
\caption{Representation of the two activation functions used in the study: the $Relu$ function (left plot) and the $Softmax$ function (right plot).}
\label{figure:Relu}
\end{figure}

Figure \ref{figure:Relu} shows shows these two functions changing with z.

In the present study, a $20 \times 20 \times 2$ neural network architecture is considered, as shown in figure \ref{figure:NN}. We have to find 440 weights and 22 biases by training the model with the labeled data set generated using MC simulation. A converged model will be able to infer the correct label with real inputs, which is to decide the type of event according to the hit information for 20 clock cycles.

\subsection{Building and training the neural network}
To build this MLP model, we use the Python library called Pytorch \cite{pytorch}.
The activation function used for the hidden layer is the $Relu$ function and the one for the output layer is the $Softmax$ function. The specific back-propagation algorithm used is a method based on the Adam algorithm \cite{adam}.
Pytorch gives already a function using this algorithm. The final parameters are the learning rate = 0.0001, $\beta = [0.9, 0.999]$, $\epsilon= 1 \times 10^{-8}$ , and the decay weight  = 0. 
\begin{figure}[ht!]
\centering
\includegraphics[width=2.5in]{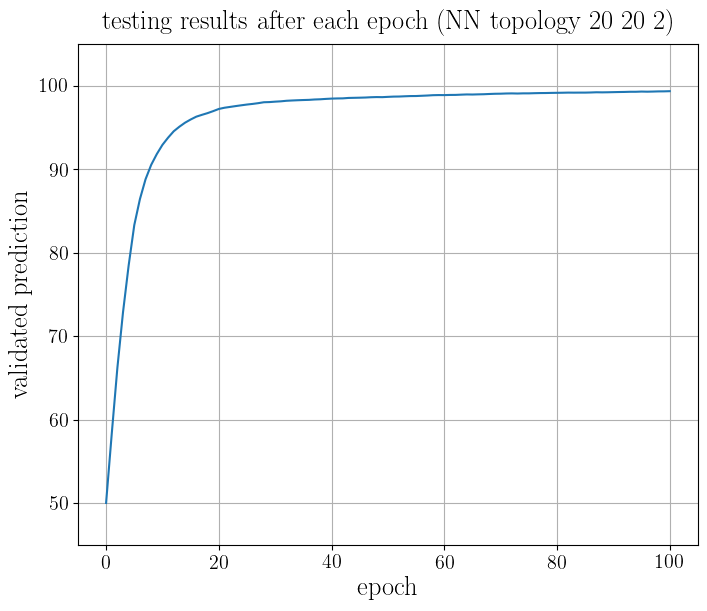}
\caption{Prediction precision on the 20 000 events data set. After 100 epochs the precision on the prediction is equal to 99\%.}
\label{figure:training}
\end{figure}

The loss function used here is the cross-entropy loss function \cite{Goodfellow-et-al-2016}. We use 180k data for the training and 20k data events for the testing sample. Figure \ref{figure:training} shows the result of the test on a data sample of 20 000 events after each epoch.
\begin{figure}[ht!]
\centering
\includegraphics[width=3.5in]{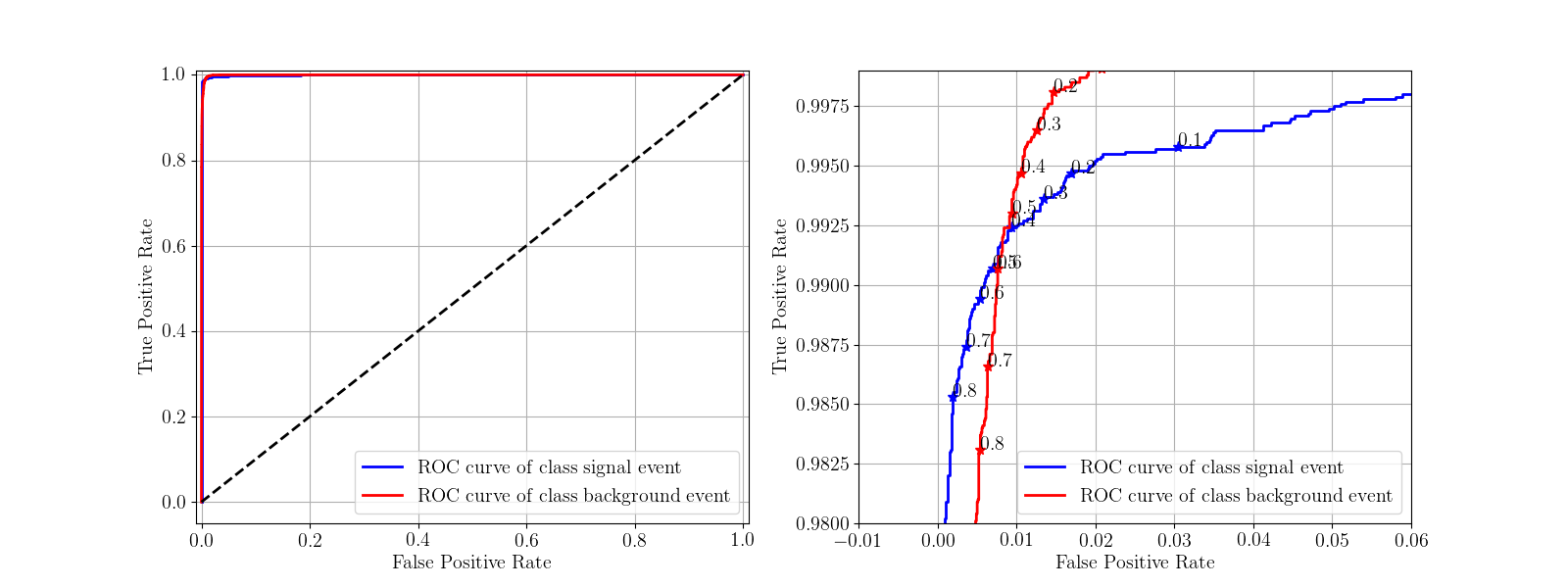}
\caption{ROC curves for the signal event class in blue and the background event class in red. The left plot shows a global view of the ROC curve. It means the classifier have good performances. The right plot is a zoom of the left plot. We have added the threshold of several point. From this, we can conclude that a threshold of 0.5 is good enough for our case.}
\label{figure:ROCcurves}
\end{figure}
 The Receiver operating characteristic (ROC) is a effective method to evaluate the performance of a classifier \cite{Bradley1997}. In ROC curves, the True positive rate ($TPR$) is considered as a function of the False positive rate ($FPR$).
 The $FPR$ is defined as:
\begin{equation}
    FPR = \dfrac{False~ Positive }{False ~Positive + True ~Positive}
\end{equation}
The $TPR$ rate is defined as:
\begin{equation}
    TPR = \dfrac{True~ Positive }{False ~Positive + True~ Positive}
\end{equation}
Figure \ref{figure:ROCcurves} shows  the ROC curves for the signal event class (in blue) and the background event class (in red).
The left plot shows a global view of the ROC curve. The right plot is a zoom of the left plot. We have added the threshold of several point. From this , we can conclude that a threshold of 0.5 is good enough for our case.

\subsection{Neural network architecture study}

\begin{figure}[ht!]
\centering
\includegraphics[width=2.9in]{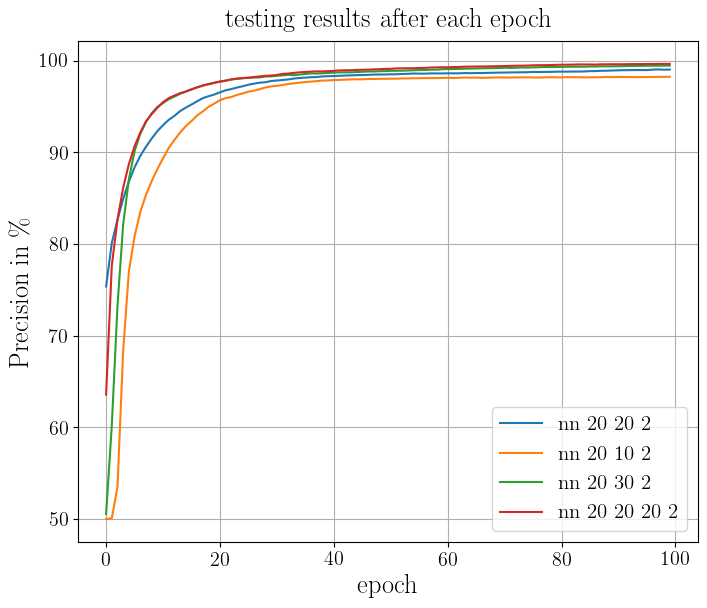}
\caption{Prediction precision on 20 000 events data set. The blue curve is for the neural network $ 20 \times 20 \times 2$, the orange curve is for the case $ 20 \times 10 \times 2$, the green curve for $ 20 \times 10 \times 2$ and the red curve for $ 20 \times 20 \times 20 \times 2$. }
\label{figure:shape_comp}
\end{figure}

Three other neural network architectures have been studied in addition to the $ 20 \times 20 \times 2$ one : the $ 20 \times 10 \times 2$, the $ 20 \times 30 \times 2$ and the $ 20 \times 20 \times 20 \times 2$ structures. Their performances are shown in figure \ref{figure:shape_comp}. Each curve represents the learning process of the neural network. All the 4 models are trained with the same data set generated with MC simulations as described in the previous chapter. The main difference between the four different curves is coming from the learning speed. We see that the more complex is the architecture of the neural network, faster it learns. Note also that the $ 20 \times 10 \times 2$ architecture has a lower prediction precision after 50 epochs of 98\% in comparison to the 99\% of the three other architectures. The models $ 20 \times 30 \times 2$ and  $ 20 \times 20 \times 20 \times 2$ achieved a precision of 99 \% faster than the $ 20 \times 20 \times 2$ model. We decided to implement the $20\times 20\times 2$ model because of its better  ratio cost in logic resources versus performance.

\subsection{Accuracy scan}

\begin{figure}[ht!]
\centering
\includegraphics[width=2.8in]{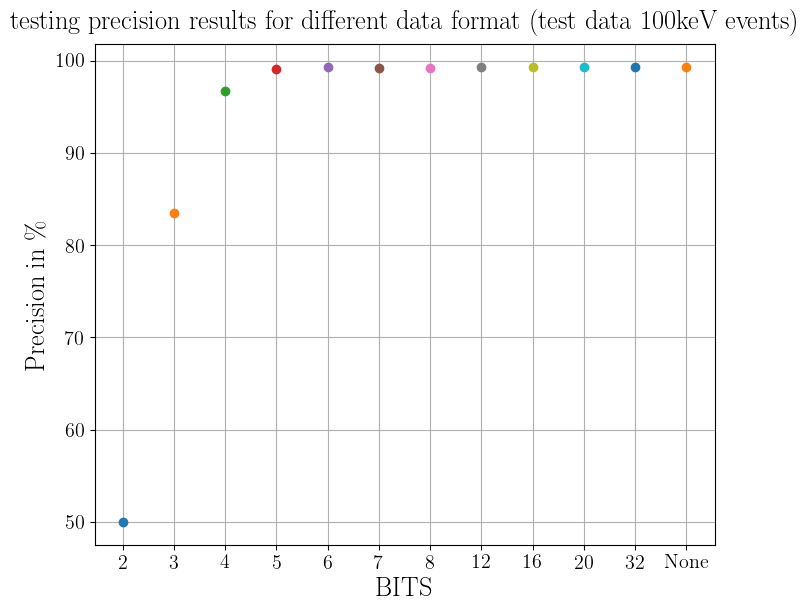}
\caption{Comparison of the neural network output as a function of the number of bits. The none value corresponds to unmodified model results. The results shows that with a encoding of the parameters on 6 bits we achieve the same precision as a neural network using parameter in double precision.}
\label{figure:bit_comp}
\end{figure}

In order to guide the firmware implementation, we need to first determine the data width, since Two’s complement is used in FPGA as a method of signed number representation \cite{DSP48E1}. We thus convert all the parameters into different bit widths of the Two’s complement and verify their accuracy. The results are shown in figure \ref{figure:bit_comp}.  
The figure shows that with 6 bits, we have the same precision as the model using floats as parameters.

\section{Firmware implementation}
\subsection{Requirements}
The basic and most important timing requirements for the level 1 trigger decision of the JUNO experiment are listed as below:
\begin{itemize}
\item Trigger decision latency less than 600 ns
\item Trigger decision running synchronized with the system clock
\end{itemize}
These requirements defined the implementation target, which is that the MLP should be able to give a trigger decision no later than 600 ns after receiving the hit information, and it should be able to give a decision every 16 ns if a 62.5 MHz system clock is used.

There are mainly 3 important steps for implementing the MLP model: the parameter and hit loading, the calculation and the verification. We use Verilog and vivado codes \cite{UG936}  to implement the firmware in order to get full control at every step.

\subsection{Parameter and hit loading}
As explained in the previous sections, the target MLP model topology is $20 \times 20 \times 2$, it has 22 neurons, each of which contains 20 weights and 1 bias; in total it includes 440 weights and 22 biases. According to MLP accuracy scan study, the use of 6 bits is accurate enough to represent all the parameters. We thus chose 12 bits Two's complement as the method of signed number representation to represent each parameter. It not only guaranties the accuracy but also facilitates the future calculation (as it will be explained in the calculation section).

A 5544 bits constant is used to keep all the parameters, the first 5040*(12*21*20) bits represent the 420 parameters for the 20 neurons in the hidden layer and the last 504*(12*21*2) bits represent the 42 parameters for the 2 neurons in the output layer.
Looking  at the distribution of the hit information from the simulated MC data, we can see that for pure dark noise coincidences, the number of fired PMTs in the 16 ns window will never exceed 20, so 8 bits Two's complement is more than enough to represent the hit information for each system clock cycle.

In principle, we need 20 clock cycles to collect the 20 inputs for the MLP, since the latency only counts from the ending of loading input until the appearing of the first trigger decision. We use a 160*500 bits BRAM to store 500 events. Each event contains 20 8-bits-wide inputs as the hit information for 20 clock cycles.

\subsection{Calculation}
From the definition of the MLP, there are mainly 3 operations that need to be implemented: multiply, accumulate, and $Relu$. Each neuron contains 20 weights and 1 bias, which correspond to 20 multiplies of of each hit and weight, and one accumulation of 20 results and 1 bias.
In order to get the best performance, all the multipliers are implemented with DSP48 core inside Kintex 7 FPGA, with pipeline stages set to 3. The multiplier can operate at  a speed higher than 200 MHz \cite{DS128}, with a fixed latency of 3 clock cycles.

\begin{figure}[b!]
\centering
\includegraphics[width=2.0in]{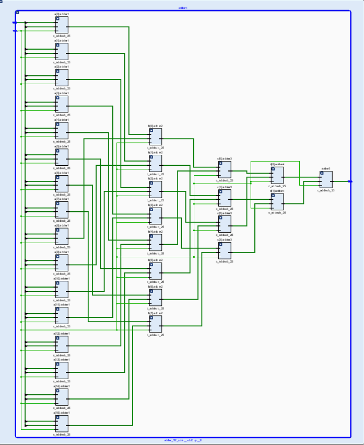}
\caption{Structure of the implementation of the accumulation operation for 32 numbers. They are first separated into 2 groups, and the 16 adders in the first level get 16 sums as the input to the 8 adders in the second level. This continues to operate until the fifth level where we get the final result.}
\label{figure:figurefirm1}
\end{figure}

A two number adder is the basic function module for the accumulation, 32 numbers are separated into 2 groups: 16 adders in first level get 16 sums as the input to the 8 adders in the second level, then 8 sums as the input to another 4 adders in the third level, until the fifth level gets the final result. Each level operates synchronized to the system clock. This implementation has the highest operation frequency while keeping the latency fixed to 5 clock cycles. The structure of the accumulation is presented in figure \ref{figure:figurefirm1}.

Since all the data are signed numbers, $Relu$ is very easy to implement by checking the highest bit. If it’s 1, it means that the value is negative and all outputs are 0. If it’s 0, the value will be transfered without any change.
There is a $Softmax$ function for the output in the MLP model. Since it’s a binary classification, by setting a cut at 0.5, we can simply output the level 1 trigger accept by checking the value of two neurons. If the neuron which represents the signal event is larger than the one which represents the dark noise event, the level 1 trigger accept is one, which means it is a signal event. Otherwise it is a dark noise event.

\begin{figure}[ht!]
\centering
\includegraphics[width=2.5in]{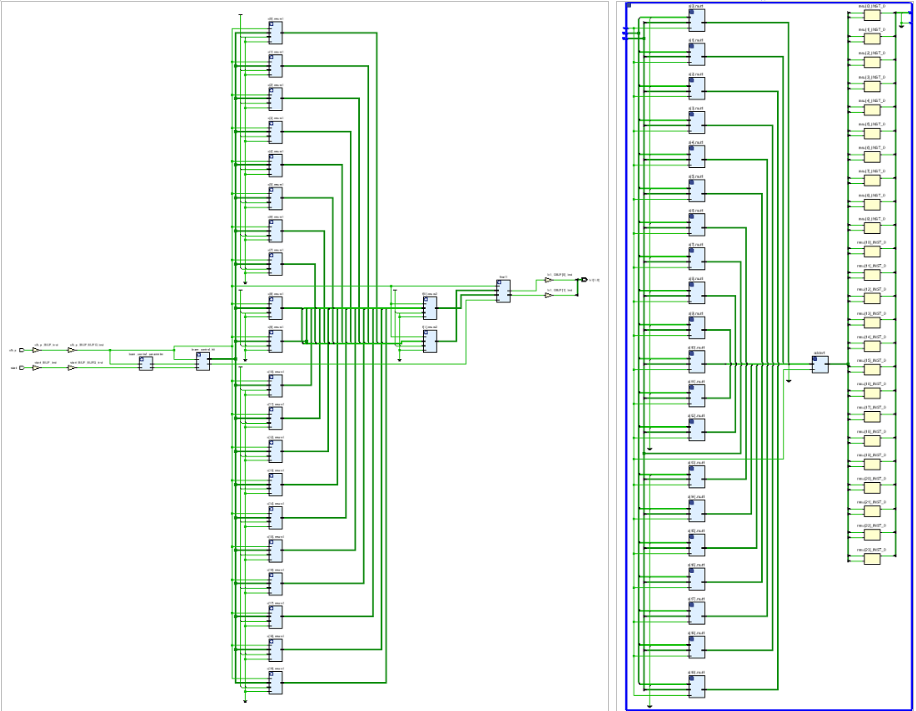}
\caption{The left figure represents the top level structure of the firmware: it includes two BRAM for parameters loading and 20 neurons in the hidden layer as well as the two neurons in the output layer, it also shows the output operation and the link to two SMA for hardware performance evaluation.
The right picture shows the detail structure of one neuron, it includes 20 multiplier implemented by DSP48 IPcore, as well as the accumulator described in figure 12 }
\label{figure:figurefirm2}
\end{figure}

The overview of the top level and one of the neuron is represented in figure \ref{figure:figurefirm2}.

\subsection{Verification}
It is important to verify the correctness of MLP operation.
Two methods are used to verify the calculation procedure. The first one is performed using simulations, with both post-implementation timing simulation and post-synthesis function simulation. We can check all the intermediate results by comparing with the software results. 
The second method uses chipscope \cite{UG936}, due to the limitation of logic resources, we are able to check all the 22 neuron’s outputs with 22 ILA probes.
These methods allow us to check that the firmware calculation is indeed correct.

\subsection{Performance}

\begin{figure}[t!]
\centering
\includegraphics[width=2.8in]{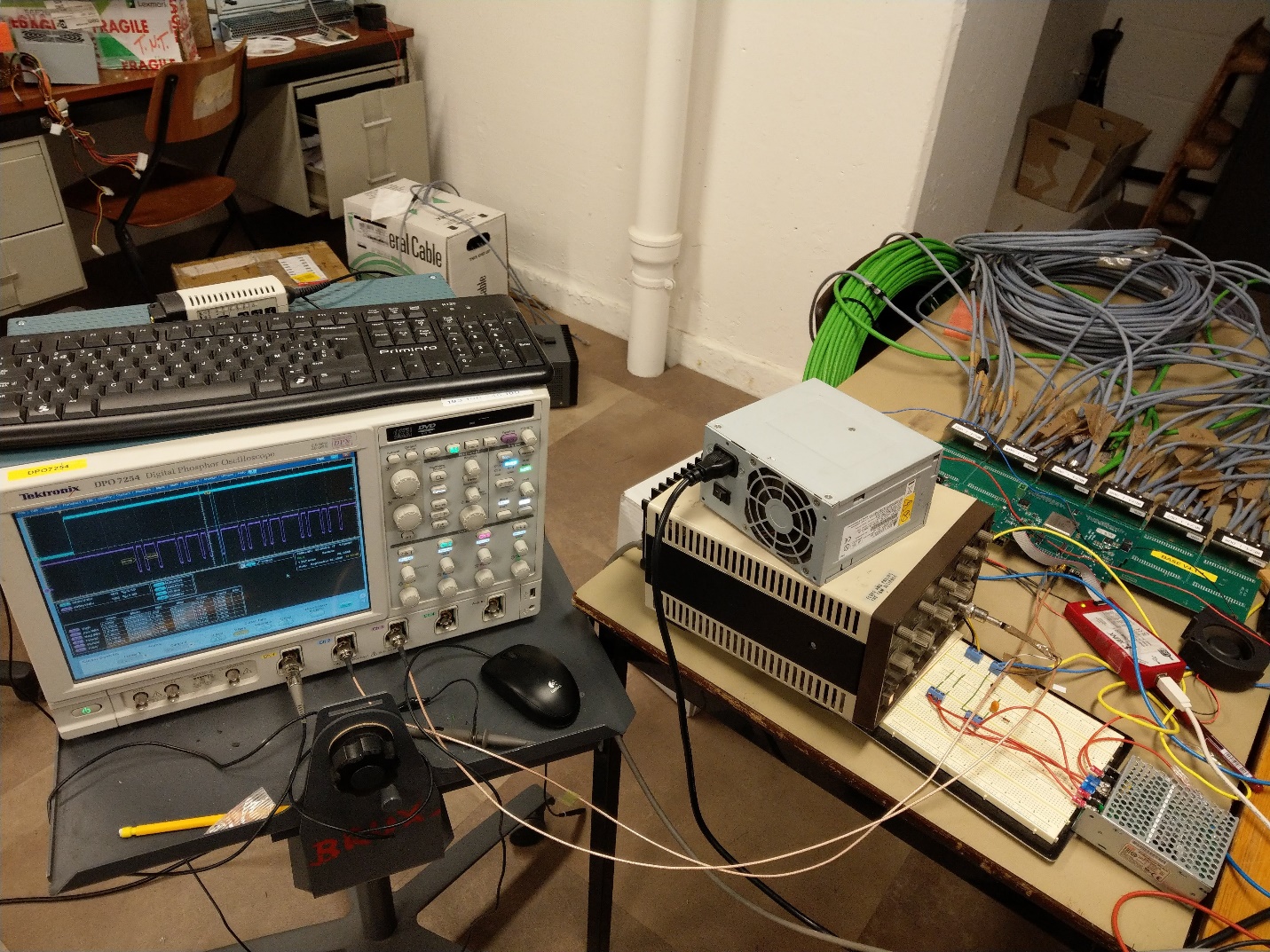}
\caption{This picture shows the hardware performance evaluation platform, on the right is the BEC with TTIM powered by redundant power supply, on the left side one can see the DPO7000 oscilloscope displaying the waveform of the two signals transferred through two SMA cables to the TTIM.  }
\label{figure:figurefirm4}
\end{figure}

\begin{figure}[t!]
\centering
\includegraphics[width=2.8in]{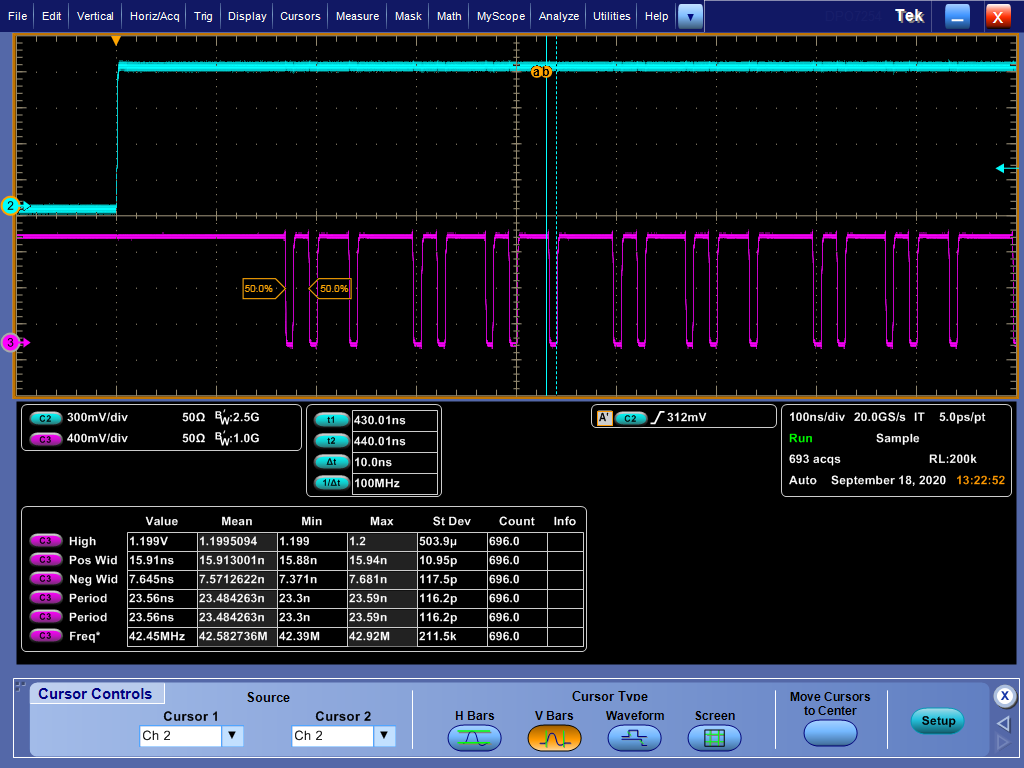}
\caption{Waveform of two SMA from the FPGA. The blue waveform is an enable signal to enable 500 inferences. The pink waveform is the results of the inferences. The oscilloscope is running at infinite persistence mode. The sharp edge of the pink signal shows that the operations are always at fixed latency. The minimum width of the pink pulse of 8 ns shows that the hardware is able to refresh the inference at a speed of 125 MHz. The 16 clock cycles difference between the enable and the first valid inference result shows that the latency is 128 ns.
}
\label{figure:figurefirm3}
\end{figure}

Besides the correctness of the operation, the timing performance is also critical. We use two spare IOs of the TTI: one is used as the enable signal which enable the operation periodically, the other one is used as the level 1 trigger accept output. During each enable cycle, there are 500 trigger decisions happening.
The hardware setup is shown in figure \ref{figure:figurefirm4}. On the right side we can see the BEC and the TTIM, powered with redundancy power supply. Two SMA cables transfer the two TTIM signals to a DPO7000 oscilloscope. On the left side. The oscilloscope is triggered with enable signal and  it is running at infinite persistence mode. A screenshot of the oscilloscope is shown in figure \ref{figure:figurefirm3}.
From the infinite persistence mode picture, the sharp edges show that the operations have few jitters. The minimum pulse width represents the current operation frequency, which is 125 MHz. The 16 clock distance between the beginning of enable signal and the first result shows that the processing latency is fixed to 128 ns.

A comparison between the performance of our implementation and the results of reference \cite{Gaikwad} is presented in table \ref{table:comparison}. The work of reference \cite{Gaikwad} implements a $7 \times 6 \times 5 $ MLP architecture in a Artix 7 35T FPGA, it uses less logic resource mainly because the MLP model is smaller, and partly because the model has been implemented by Xilinx system generator design toolbox, not all the neurons are implemented by DSP48 hardcore. On the other hand, our implementation have shorter latency which is critical for level 1 trigger decision.

\begin{table}[ht!]
\centering
\caption{Comparison between the results obtained in the work of reference \cite{Gaikwad} (left column) and the results of this work (right column).}
\begin{tabular}{|l|l|l|}
\hline
                  & Reference \cite{Gaikwad}   & This work     \\ \hline
MLP topology      & $7 \times 6 \times 5 $      &$ 20\times 20 \times 2  $     \\ \hline
FPGA model        & Artix 7 35T & Kintex 7 325T \\ \hline
Data width        & 16          & 12            \\ \hline
Latency           & 270 ns      & 128 ns        \\ \hline
Power consumption & 241 mW      &        350 mW \\ \hline
BRAM              & 0           & 0             \\ \hline
DSP               & 81          & 440           \\ \hline
LUT               & 3466        & 16563         \\ \hline
\end{tabular}
\vspace*{0.2cm}

\label{table:comparison}
\end{table}

\section{Conclusions}
This study shows an efficient cooperation between 3 different specialized subjects, and defines a smooth working flow. Each stage is implementation oriented. The use of simulated data as a labeled dataset links machine learning with physics. The implementation of the MLP model in FPGA makes the trigger decision algorithm flexible and easy to maintain: by simply changing the parameter constants in the firmware it is very easy and efficient to upgrade the model for future needs.

The design takes full advantage of the parallelism of FPGA and the high speed operation of the embedded DSP hardcore. The model has been implemented in a Kintex 7 FPGA, with a trigger latency of 128 ns at a clock frequency of 125 MHz. It fulfils the timing requirements from the level 1 trigger decision and indicates the feasibility of using MLP for level 1 trigger decision in the  JUNO experiment. It also gives the limitation from the FPGA to the model, which is the amount of DSP48. Once a model has been trained, it is easy to find the most suitable FPGA according to the number of layers and the number of neurons in each layer.

In the future, we will make a more realistic signal simulation. We will generate signal events in a spectrum of energy range, and we will generate a bigger dataset to optimize the MLP model. Following the development procedure presented here, a better performance is foreseen. Since the current trigger algorithm based on vertex fitting does not require any use of DSP48, it would be very interesting to put the MLP model into the current CTU hardware, and to verify the efficiency of the new proposed trigger during real data taking.

\bibliographystyle{unsrt}
\bibliography{tns}

\end{document}